\documentclass{vietnam}

\usepackage{amsmath}
\usepackage{xspace}

\newcommand{\GeV}{\ensuremath{\,\mathrm{GeV}}\xspace}

\newcommand{\fb}{\ensuremath{\,\mathrm{fb}}\xspace}

\newcommand{\order}[1]{\mathcal{O}\!\left(#1\right)}

\newcommand{\eq}[1]{Eq.~(\ref{#1})}
\newcommand{\bib}[1]{Ref.~\cite{#1}}
\newcommand{\fig}[1]{Fig.~\ref{#1}}

\newcommand{\be}{\begin{equation}}
\newcommand{\ee}{\end{equation}}
\newcommand{\bea}{\begin{eqnarray}}
\newcommand{\eea}{\end{eqnarray}}
\newcommand{\beqn}{\begin{eqnarray}}
\newcommand{\eeqn}{\end{eqnarray}}
\newcommand{\bal}{\begin{align}}
\newcommand{\eal}{\end{align}}
\newcommand{\bitem}{\begin{itemize}}
\newcommand{\eitem}{\end{itemize}}

\newcommand{\crn}{\nonumber \\}

\newcommand{\Photo}{}

\begin{document}
\vspace*{0.1cm}
\rightline{FTUV-13-1016,IFIC-13-75,KA-TP-32-2013,LPN13-076,SFB/CPP-13-76}

\vspace*{3.9cm}
\title{NLO QCD CORRECTIONS TO $WZJJ$ PRODUCTION AT THE LHC}

\author{Francisco~Campanario$^1$, Matthias~Kerner$^2$, Le~Duc~Ninh$^2$\footnote{Speaker} and Dieter~Zeppenfeld$^2$}
\address{$^1$Theory Division, IFIC, University of Valencia-CSIC, E-46980
  Paterna, Valencia, Spain.\\
  $^2$Institute for Theoretical Physics, KIT, 76128 Karlsruhe, Germany.}

\maketitle\abstracts{
We present a summary of the first calculation of NLO QCD corrections to 
$W^\pm Zjj$ production with leptonic decays at the LHC. 
Our results show that the next-to-leading order corrections reduce
significantly the scale uncertainties.}

\section{Introduction}

The study of di-boson production in association with two jets at the LHC 
is important not only as a background to many physic searches, but also, as a
signal since it is sensitive to 
the quartic gauge couplings of the Standard Model (SM) and 
the four-vector-boson scatterings 
of the type $VV\to VV$ where the initial gauge bosons are radiated from the 
incoming (anti-)quarks. At leading order (LO), there are three different 
production mechanisms. The vector boson fusion mechanism of 
the order $\order{\alpha^6}$ includes in particular the electroweak (EW) gauge-boson
scattering and it has been studied at NLO QCD in Refs.~\cite{Jager:2006zc,Jager:2006cp,Jager:2009xx,Denner:2012dz,Campanario:2013eta}.  
In addition, the production of three EW gauge bosons, with one off-shell
gauge boson decaying into a quark-antiquark pair, is a second source of
$VVjj$ events at order $\order{\alpha^6}$ and will be available at NLO QCD with
leptonic decays via the {\texttt{VBFNLO}} program~\cite{Arnold:2008rz,Arnold:2012xn}.

Finally, there are QCD contributions of the order $\order{\alpha_s^2\alpha^4}$. 
The NLO QCD corrections to this mechanism have been calculated 
for the $W^+W^-jj$ production in Refs.~\cite{Melia:2011dw,Greiner:2012im}, 
for the $W^+W^+jj$ case in \bib{Melia:2010bm} and recently in
\bib{Gehrmann:2013bga} for the $\gamma \gamma jj$ and in \bib{Campanario:2013qba} 
for the $W^\pm Zjj$ channels.    
Indeed, the last processes with one undetected lepton are the main backgrounds to the 
same-charge $W^+W^+jj$/$W^-W^-jj$ observation at the LHC. 

Since the above three production modes
peak in different regions of phase space, and because of their largely
orthogonal color structures, interference effects between these modes are generally
unimportant and can be neglected in most applications.

In the following, we consider the QCD induced $W^\pm Zjj$ production modes within the SM. 

\section{Calculation}
\begin{figure}[h]
  \centering
\includegraphics[width=0.43\textwidth]{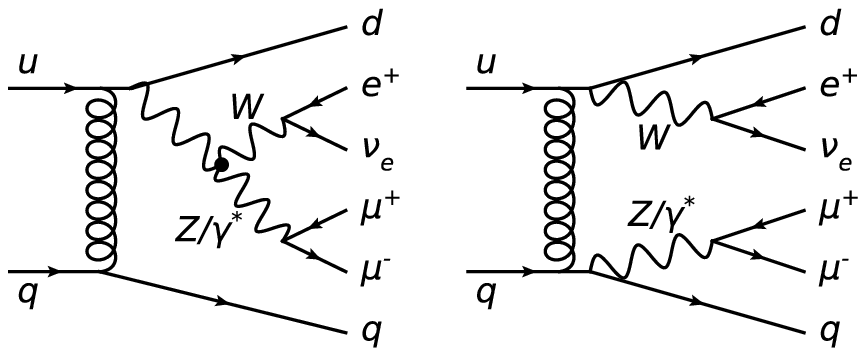}
\includegraphics[width=0.43\textwidth]{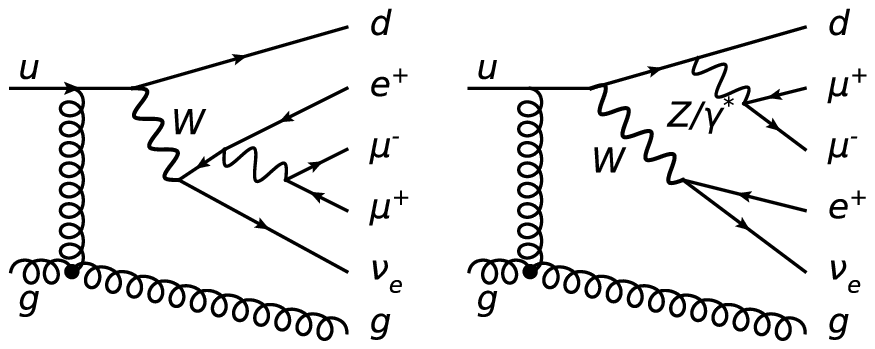}
\caption{Representative tree-level Feynman diagrams.}
\label{fig:feynTree}
\end{figure}

In this section, we define the problem and summarize our calculational method. 
The processes are
\bea
pp \to e^{+} \nu_e \mu^{+} \mu^{-} jj + X,\crn
pp \to e^{-} \bar{\nu}_e \mu^{+} \mu^{-} jj + X,
\label{eq:process}
\eea
where $p,j=g,u,d,c,s$ and the corresponding anti-quarks. 
The subprocesses with $t,b$ are neglected since their contributions 
are very small. At LO, only the 
QCD mechanism as illustrated in \fig{fig:feynTree} is included.  
The dominant contribution comes from the diagrams where 
both $W^\pm$ and $Z$ can be simultaneously on-shell. 
This is why we refer to the processes \eq{eq:process} as $W^\pm Zjj$ production. 
However, the sub-dominant diagrams with one resonating 
gauge boson are also included, hence the total amplitudes 
are QCD and EW gauge invariant. The challenge is then to calculate the NLO 
QCD corrections to get theoretical prediction at order $\order{\alpha_s^3\alpha^4}$. 

\begin{figure}[h]
  \centering
\includegraphics[width=0.4\textwidth]{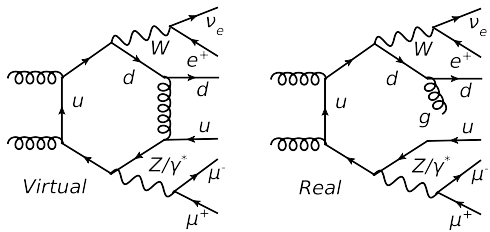}
\caption{Representative virtual and real-gluon emission Feynman diagrams.}
\label{fig:feyn_virt_real}
\end{figure}

At NLO, there are the virtual and the real 
corrections as shown in \fig{fig:feyn_virt_real}. 
There are $90$ and $146$ subprocesses for the LO and real-emission contributions, respectively.   
Both the virtual and real corrections are, apart from the 
UV divergences in the virtual amplitude which are removed by the renormalization of $\alpha_s$, 
separately infrared divergent. These divergences cancel in the sum for 
infrared-safe observables such as the inclusive cross section and jet distributions. 
We use the dimensional regularization method~\cite{'tHooft:1972fi} 
to regularize the UV and the infrared divergences and apply the 
Catani-Seymour dipole subtraction algorithm~\cite{Catani:1996vz} to combine the virtual 
and the real contributions. 
The most difficult part 
of the calculation is computing the $2$-quark-$2$-gluon virtual
amplitudes with up to six-point rank-five one-loop 
tensor integrals. There are $84$ six-point diagrams for each of seven independent subprocesses. 
The $4$-quark group is much easier with only $12$ hexagons for the most complicated subprocesses 
with same-generation quarks. Given the complexity of the calculation, we have 
implemented two independent codes. 
Details of the implementation and cross checks are given in \bib{Campanario:2013qba}.

\section{Numerical results}
As input parameters, we choose the following inclusive cuts. 
For leptons: 
\bea
 p_{T,\ell}\ge 20 \GeV,\;\;\; |y_{\ell}|\le 2.5,\;\;\; E_{T,\text{miss}} \ge 30 \GeV,\;\;\; m_{\ell^+\ell^-} \ge 15 \GeV,
\eea
where the last cut is for any pair of 
opposite-charge leptons. For jets, we use the anti-$k_t$ algorithm~\cite{Cacciari:2008gp} with radius $R=0.4$ and 
\bea
p_{T,\text{jet}}\ge 20\GeV,\;\;\; |y_{\text{jet}}|\le 4.5.
\eea
We also impose a requirement on the lepton-lepton and lepton-jet
separation in the azimuthal angle-rapidity plane $\Delta R_{\ell(\ell,j)} \ge 0.4$, where 
only jets passing the above cuts are considered. 
As the central value for the factorization and renormalization scales, we choose
\bea
\mu_{F}=\mu_{R}=\mu_{0}=
\left(\sum_\text{jet}  p_{T,\text{jet}} + 
\sqrt{p_{T,W}^2+m_W^2} +
\sqrt{p_{T,Z}^2+m_Z^2}\right)/2,
\eea
where $p_{T,V}$ and $m_V$ with $V$ being $W$ or $Z$ are the reconstructed
transverse momenta and invariant masses of the decaying bosons and the sum 
includes only jets passing all cuts.
\begin{figure}[t]
  \centering
  \includegraphics[width=0.45\textwidth]{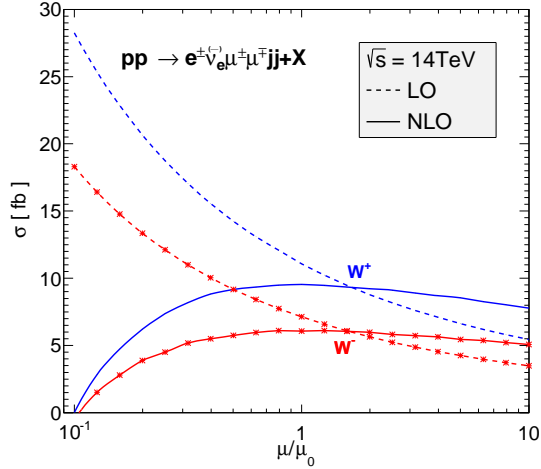}
\caption{Scale dependence of the LO and NLO cross sections at the LHC. 
The curves with and without stars are for $W^-Zjj$ and $W^+Zjj$ productions, 
respectively.}
\label{fig:scale}
\end{figure}

In \fig{fig:scale}, we plot the cross section calculated at LO and NLO as functions 
of $\mu = \mu_{F} = \mu_{R}$. 
\begin{figure}[t]
  \centering
  \includegraphics[width=0.43\textwidth]{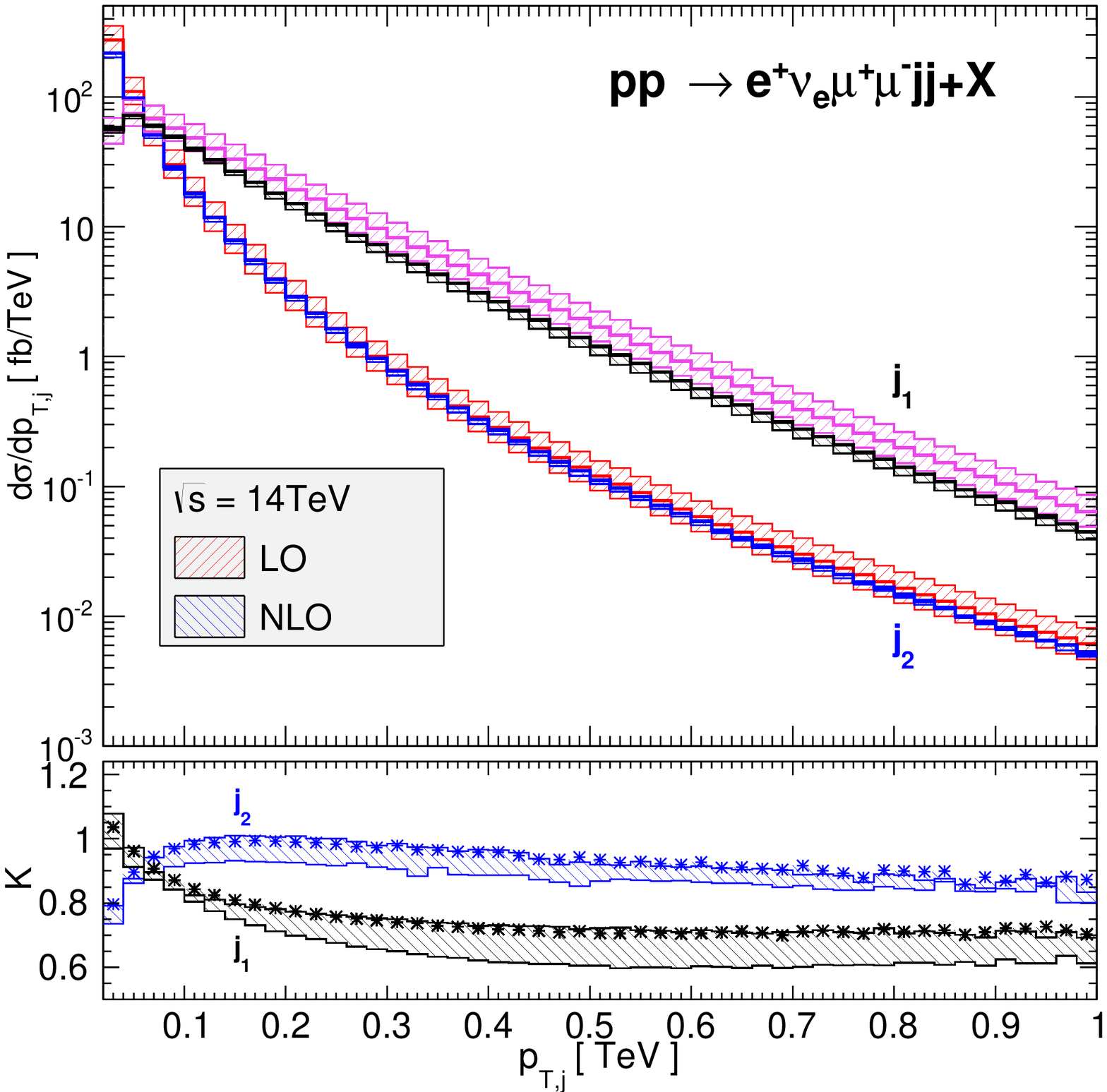}
  \includegraphics[width=0.43\textwidth]{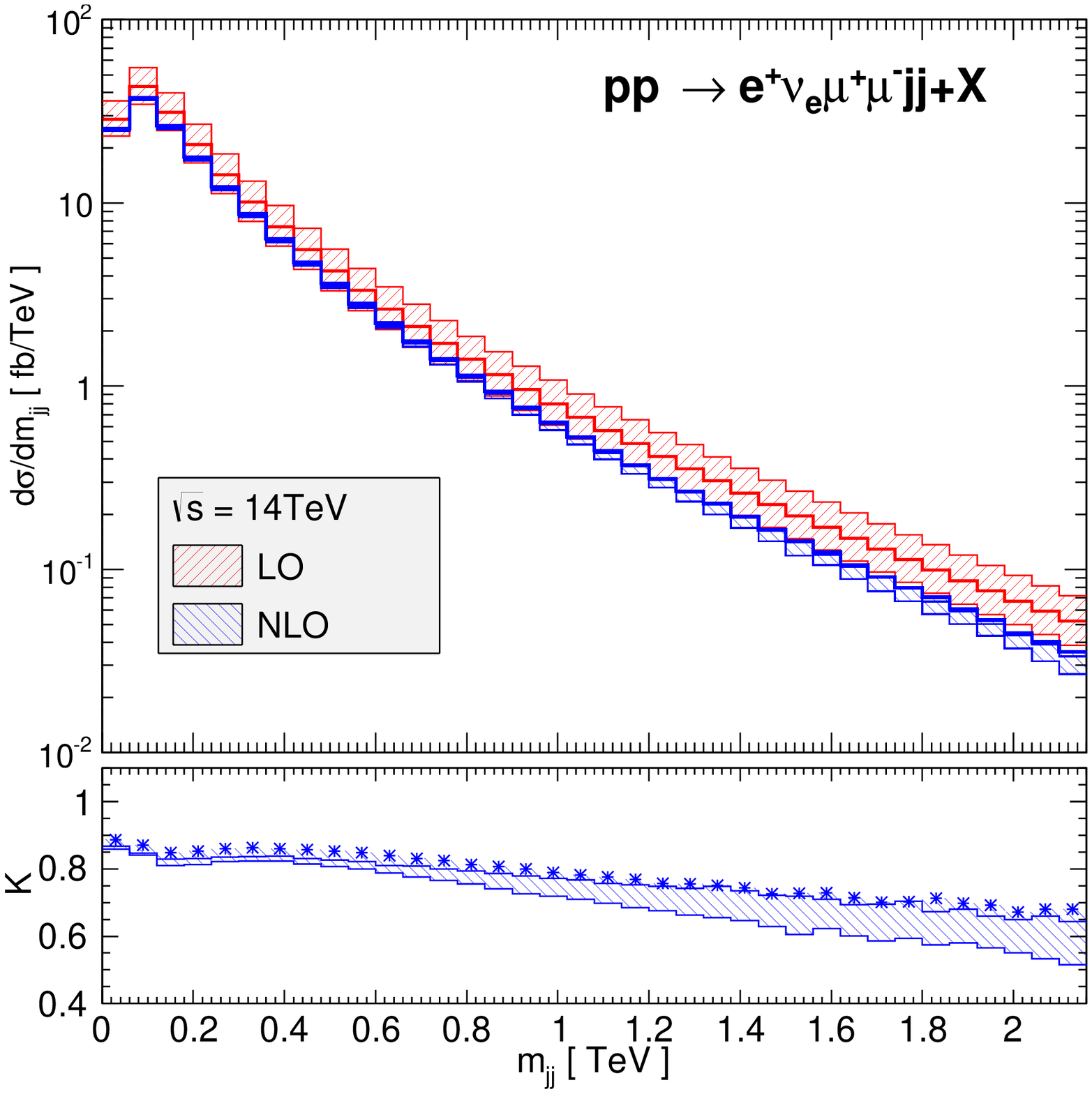}
  \caption{Differential cross sections and K-factors for the transverse momenta (left) and 
the invariant mass (right) of the two hardest jets. 
The bands describe $\mu_0/2 \le \mu_F=\mu_R\le 2\mu_0$ variations. 
The $K$-factor bands are due to the scale variations of the NLO results, 
with respect to $\sigma_\text{LO}(\mu_0)$. 
The curves with stars in the lower panels are for the central scale, while the two solid lines correspond to $\mu_F = \mu_R=2\mu_0$ 
and $\mu_0/2$.
}
\label{fig:dist_jet}
\end{figure}
As expected, 
we observe a significant reduction in the scale dependence around the central value 
$\mu_0$ when the NLO contribution is included. For both $W^+$ and $W^-$ cases, the uncertainties 
obtained by varying $\mu_{F,R}$ by factors $1/2$ and $2$ around the central value 
are $50\%$ at LO and $5\%$ at NLO. 
At $\mu = \mu_0$, we get $\sigma_\text{LO}=11.1^{+3.2}_{-2.3}\fb$($7.1^{+2.0}_{-1.5}\fb$) 
and $\sigma_\text{NLO}=9.5^{+0.0}_{-0.4}\fb$($6.1^{+0.0}_{-0.3}\fb$) for the $W^+$($W^-$) case. 
 By varying the two scales separately, we observe a small dependence on $\mu_F$, while 
the $\mu_R$ dependence is similar to the behavior shown in \fig{fig:scale}. 

The distributions of the 
transverse momenta and 
the invariant mass of the two hardest jets are shown in \fig{fig:dist_jet}. 
The $K$-factors, defined as the ratio of the NLO to the LO results, are 
shown in the lower panels. The distributions at NLO are much less sensitive
to the variation of the scales than at LO. The $K$-factors vary from $0.6$ to $1$ 
in a large energy range. We have also studied a fixed scale choice such as $\mu_0^\text{fix} = 400\GeV$ and found that 
the NLO inclusive cross section as a function of the scales is stable around $\mu_0^\text{fix}$ 
and is close to the LO one as well as the dynamic scale prediction. 
However, the 
transverse momentum and the invariant mass distributions become unstable at large $p_{T}$, with 
very small $K$-factors. This is because the bulk of the inclusive cross section comes from the low energy regime 
as shown in \fig{fig:dist_jet}, but a fixed energy scale is not appropriate for all energy regimes.   
The steep increase of the $K$-factor for the transverse momentum distribution of the second hardest jet 
near $20\GeV$ is probably a threshold effect: the phase space for three-visible-jet events is opened up 
as $p_{T,j_2}$ grows well above the cut of $20$GeV. 

\section{Conclusions}
In this talk, we have reported on the first calculation of $W^{\pm}Zjj + X$
production at order $\order{\alpha_s^3 \alpha^4}$ and found K-factors close to
one. This is a part of our project to include NLO QCD corrections to 
$VVjj$ production processes at the LHC in 
the {\texttt{VBFNLO}} program~\cite{Arnold:2008rz,Arnold:2012xn}.

\section*{Acknowledgments}
LDN would like to thank the organizers, in particular Tran Thanh Van, for 
the nice conference. We acknowledge the support from the Deutsche Forschungsgemeinschaft
via the Sonderforschungsbereich/Transregio SFB/TR-9 Computational Particle Physics.
FC is funded by a Marie Curie fellowship (PIEF-GA-2011-298960) and partially by MINECO (FPA2011-23596) and by LHCPhenonet (PITN-GA-2010-264564). 
MK is supported by the Graduiertenkolleg 1694 ``Elementarteilchenphysik bei h\"ochster Energie und h\"ochster Pr\"azision''.

\end{document}